\documentclass[a4paper,11pt]{article}

\input{preamble}

\title{Commissioning of the 2.6\,m tall two-phase xenon time projection chamber of Xenoscope}

\author{M.~Adrover\orcidlink{0009-0009-4903-4125},}
\author{M.~Babicz\orcidlink{0000-0002-1017-5440},}
\author{L.~Baudis\orcidlink{0000-0003-4710-1768},}
\author[1]{Y.~Biondi\orcidlink{0000-0002-9323-0846},}
\note{Now at Karlsruhe Institute of Technology}
\author[*]{A.~Bismark\orcidlink{0000-0002-0574-4303},}
\note[*]{Corresponding author}
\author{C.~Capelli\orcidlink{0000-0003-3330-621X},}
\author[*]{A.~P.~Cimental Chávez\orcidlink{0009-0004-9605-5985},}
\author{J.~J.~Cuenca-García\orcidlink{0000-0002-3869-7398},}
\author[2]{M.~Galloway\orcidlink{0000-0002-8323-9564},}
\note{Now at University of Bern}
\author[3]{F.~Girard\orcidlink{0000-0003-0537-6296},}
\note{Now at Laboratoire de Physique Nucléaire et des Hautes Énergies}
\author{F.~J\"org\orcidlink{0000-0003-1719-3294},}
\author{S.~Ouahada\orcidlink{0009-0007-4161-1907},}
\author[*]{R.~Peres\orcidlink{0000-0001-5243-2268},}
\author{F.~Piastra\orcidlink{0000-0001-8848-5089},}
\author{M.~Rajado Silva\orcidlink{0000-0002-7663-2915},}
\author{D.~Ramírez~García\orcidlink{0000-0002-5896-2697},}
\author{C.~Wittweg\orcidlink{0000-0001-8494-740X}}
\affiliation{Physik-Institut, Universit\"at Z\"urich, 8057  Z\"urich, Switzerland}

\emailAdd{alexander.bismark@physik.uzh.ch, paloma.cimentalchavez@physik.uzh.ch, ricardo.peres@physik.uzh.ch}

\abstract{Xenoscope is a demonstrator for a next-generation xenon-based observatory for astroparticle physics, as proposed by the XLZD (XENON-LUX-ZEPLIN-DARWIN) collaboration. It houses a \qty{2.6}{m} tall, two-phase xenon time projection chamber (TPC), in a cryostat filled with \qty{\ca 360}{kg} of liquid xenon. The main goals of the facility are to demonstrate electron drift in liquid xenon over this distance, to measure the electron cloud transversal and longitudinal diffusion, as well as the optical properties of the medium. In this work, we describe in detail the construction and commissioning of the TPC and report on the observation of light and charge signals with cosmic muons.}


\hypersetup{%
  pdftitle = {Commissioning of the 2.6 m tall two-phase xenon time projection chamber of Xenoscope},
}

\begin{document}
\maketitle
\flushbottom
    \section{Introduction} 
\label{sec:intro}

Two-phase (liquid and gas) xenon time projection chambers (TPCs) are powerful detectors in the fields of astroparticle physics and rare-event searches~\cite{Baudis:2023pzu,ParticleDataGroup:2024cfk}. Ongoing experiments based on this technology have reached the multi-tonne scale and ultra-low background rates~\cite{LZ:2022ysc,PandaX-4T:2021bab,XENON:2024wpa}. Next-generation detectors plan to further increase the xenon target mass and thus the linear dimensions of the TPCs, in order to house tens of tonnes of liquid xenon (LXe)~\cite{DARWIN:2016hyl,PandaX:2024oxq,Baudis:2024jnk, XLZD:2024gxx}. At the same time, the backgrounds will be reduced such that interaction rates will be dominated by cosmic and solar neutrinos~\cite{DARWIN:2020bnc,XENON:2024ijk, PandaX:2024muv} and by second-order weak decays~\cite{XENON:2022evz}. The XENON-LUX-ZEPLIN-DARWIN (XLZD) experiment will operate a TPC with an active mass of \qtyrange{60}{80}{t}~\cite{Baudis:2024jnk, XLZD:2024gxx}. This scale-up imposes a series of technological challenges to be addressed with complementary large-scale demonstrators, such as PANCAKE~\cite{Brown:2023vgf} and Xenoscope~\cite{Baudis:2021ipf}. The PANCAKE facility can house a TPC with a \qty{2.6}{m} diameter, while Xenoscope was constructed to operate a \qty{2.6}{m} tall TPC. Both demonstrators have a capacity of \qty{\ca 400}{kg} of liquid xenon but aim to address different scale-up challenges. The design and construction of Xenoscope, without the TPC, are detailed in ref.~\cite{Baudis:2021ipf}, and first results with a \qty{53}{cm} tall purity monitor were published in ref.~\cite{Baudis:2023ywo}. 

In this work, the focus is on the new systems, in particular the \qty{2.6}{m} tall TPC, and on its commissioning phases with gaseous xenon (GXe) and LXe. We also show first scintillation and electroluminescence signals acquired from cosmic muons. The vertical demonstrator will be key to prepare the next-generation LXe observatory by testing electron drift over unprecedented distances and performing electron diffusion measurements. Furthermore, the optical properties of liquid xenon, such as the Rayleigh scattering length and vacuum ultraviolet (VUV) light absorption on impurities of the medium, are to be studied under controlled conditions at a large scale.

The paper is organised as follows: \autoref{sec:facility} gives an overview of the entire facility, \autoref{sec:tpc} details the \qty{2.6}{m} tall TPC and the systems designed and installed for its operation,  \autoref{sec:commissioning} describes the commissioning of the different systems,  the first run of Xenoscope with the dual-phase TPC and the data acquired with cosmic muons. \Autoref{sec:outlook} provides  a summary of the results and an outlook towards the next steps of the project. Additionally, a new gas-extraction line was added and commissioned during the run and is reported in \autoref{sec:appendix}.

\section{The Xenoscope facility}
\label{sec:facility}
Xenoscope was designed to support a xenon dual-phase TPC with continuous xenon purification for the demonstration of electron drift at the scale of the XLZD experiment. The physics program is divided into two phases with different inner detectors: First, the xenon purification capability was demonstrated and electron transport properties in LXe were measured using a \qty{53}{cm} tall purity monitor~\cite{Baudis:2023ywo}. In the second phase discussed in this paper, a xenon dual-phase TPC will demonstrate electron drift over \qty{2.6}{m}. Here, we give a brief overview of the cryogenic system, xenon gas handling and purification, xenon storage and recovery systems, and slow control.

The cryostat can house up to \qty{400}{kg} of LXe with the TPC requiring \qty{\ca 360}{kg}. All electrical connections, apart from the cathode high-voltage of the TPC, as well as the gas system connections are made on the top flange of the cryostat. The top flange also connects the cryostat to the support structure where it can be levelled using an in-house designed and constructed levelling system~\cite{legspatent_ch, legspatent_de}. The xenon inventory is stored in a rack of nine aluminium bottles from which it can be introduced into the cryostat via the gas system. The xenon liquefaction and thermal stability of the system during filling and operation are maintained by a cooling tower equipped with an Iwatani PC-150 pulse tube refrigerator and a secondary $\text{LN}_2$ cooling system. During operation, the xenon inventory is purified in gaseous phase through a hot metal getter via continuous recirculation with a double-diaphragm compressor, designed for flows up to \qty{100}{slpm}. A custom-made heat exchanger minimises heat losses from cold outgoing and warm incoming gaseous xenon during recirculation~\cite{Baudis:2021ipf}. An $\text{LN}_2$ cooling jacket, the precooler, which is attached to the top section of the inner cryostat allows for a more rapid cooldown of the system before filling and provides additional cooling in the process. After operation, or in case of an emergency, the LXe inventory can be recovered by cryogenic pumping of xenon gas to the bottle rack or directly in liquid form with a gravity-assisted liquid recuperation and storage system~\cite{Baudis:2023ywo}.

Continuous control and monitoring of the experiment are ensured by a slow control system developed in-house. The system uses a micro-service architecture based almost exclusively on open-source software. It consists of several independent subsystems that take care of data collection, data ingestion and visualisation, control, monitoring, alarming, and orchestration of the services. The system is designed with redundant hardware and network connections while the micro-service architecture makes sure that the failure of one software service does not affect others.

\section{The time projection chamber}
\label{sec:tpc}

When an interacting particle deposits energy via scattering with a xenon atom, it induces the production of scintillation photons, ionisation electrons and heat. The latter is undetected for this detection principle. Photosensors, placed above the target in Xenoscope, detect the VUV scintillation photons as a prompt signal, denoted S1. A drift field between the cathode and the gate electrodes prevents the recombination of the ionisation electrons, which are drifted towards the gas-liquid interface, located between the gate and anode electrodes. The stronger field between these electrodes extracts the electrons into the gas phase, where electroluminescence occurs. The delayed VUV photons created are also detected by the photosensors, giving rise to the S2 signal. From the S1 and S2 signals of an event, its deposited energy, position and type of recoil of the interaction can be inferred. Further details and illustration of the TPC concept can be found in ref.~\cite{Baudis:2023pzu}.

In Xenoscope, the TPC is contained within the inner pressure vessel of the double-walled stainless steel cryostat, designed for nominal operation at \qty{2}{bar} and \qty{177}{K}. It consists of a field cage, a set of three electrodes (cathode at the bottom, gate and anode at the top), and one photosensor plane at the top, above the anode. The TPC is off-centre with respect to the inner vessel and the liquid-level monitoring and setting system is placed on the major clearance side.

In this section, we describe the \qty{2.6}{m} tall TPC and its subsystems in detail. The field cage and its mechanical assembly are detailed in \autoref{sec:cage}, followed by the high-voltage (HV) delivery system in \autoref{sec:hv}, the liquid level monitor and control systems in \autoref{sec:level}, and the description of the top photosensor plane in \autoref{sec:sipm}. 
    \subsection{The TPC field cage}\label{sec:cage}

The field cage of the TPC maintains the homogeneity and strength of the electric field along the full vertical dimension by creating a vertical succession of equipotential planes. In Xenoscope, the field cage is built using 173 oxygen-free high thermal conductivity (OFHC) copper rings with an internal diameter of \SI{16}{cm}. In order to bias the copper rings and supply a gradual voltage gradient, they are interconnected by a dual-resistor chain. For the assembly, we have used CRHV2512AF1G00FKET $\SI{1}{G\Omega}$ (with $\SI{1}{\percent}$ tolerance) resistors that are mounted on printed circuit boards (PCBs) and connected in parallel between two adjacent rings. The total power dissipation of the entire resistor chain is \SI{\ca 36}{mW}.

The \SI{2.6}{m} field cage has a modular structure, consisting of five \SI{52}{cm} modules. In each module, the copper rings are mounted on six pillars made of polyamide-imide (PAI) and secured on the inside with polytetrafluoroethylene (PTFE) parts. The use of PAI in a large-scale TPC is a novelty, motivated by its mechanical properties against thermal contraction, as well as its high dielectric strength and low outgassing rate compared to other polymers  \cite{outgassingTorlon}.

   \begin{figure}[!htb]
\begin{tabular}{cc}
    \begin{minipage}{0.5\textwidth} \hspace{-0.22cm} \includegraphics[height=11cm, width=1.02\textwidth]{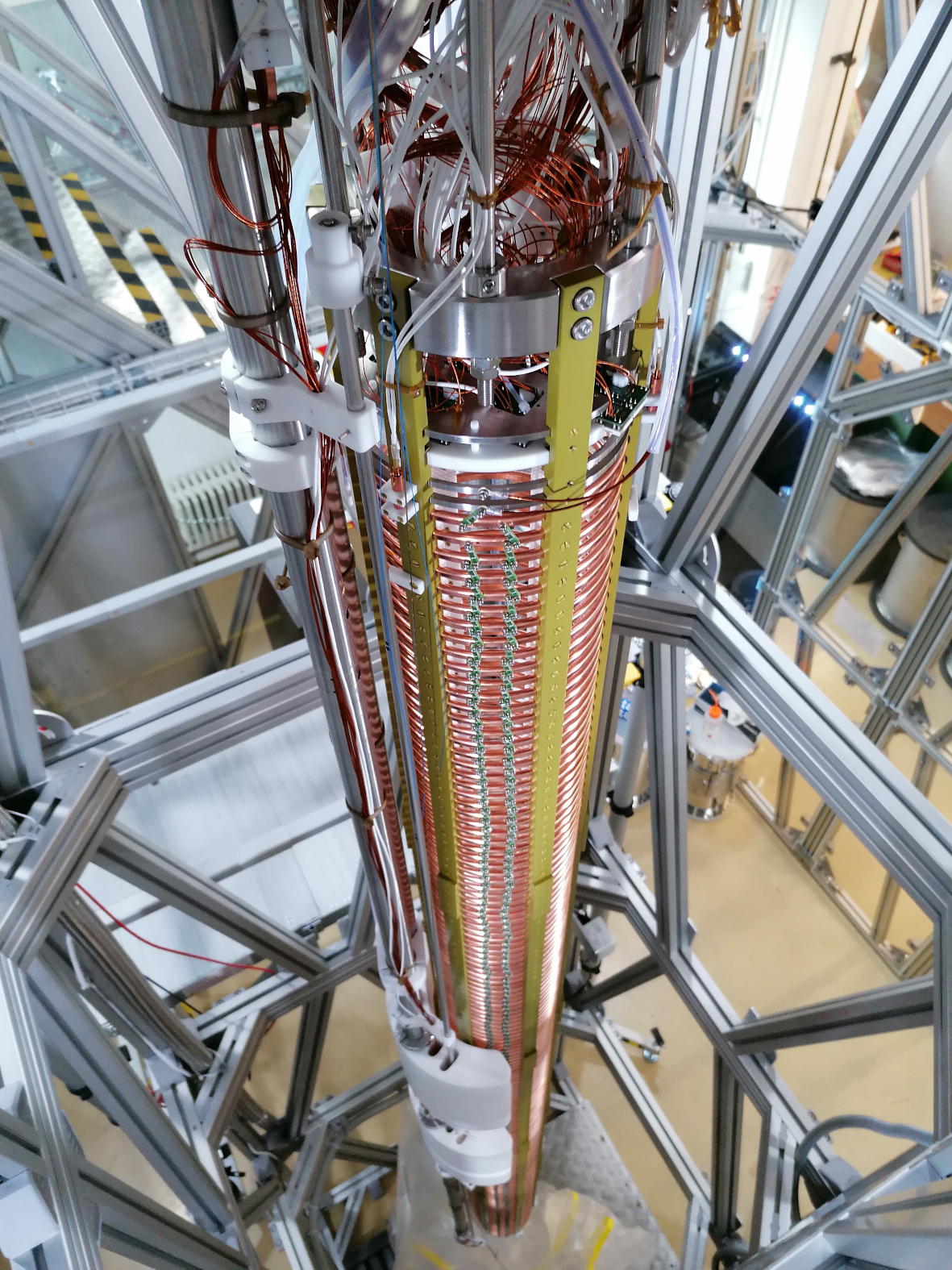} \end{minipage} & \hfill \begin{minipage}{0.5\textwidth} \includegraphics[trim={0 0 0 0},clip, width=0.9\textwidth]{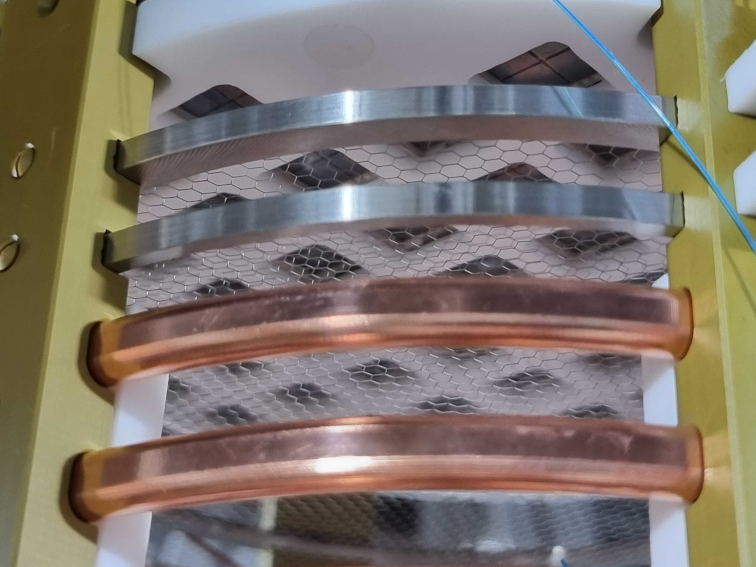} \\ \vfill \includegraphics[trim={0 2.3cm 0 2.7cm}, clip, width=0.9\textwidth]{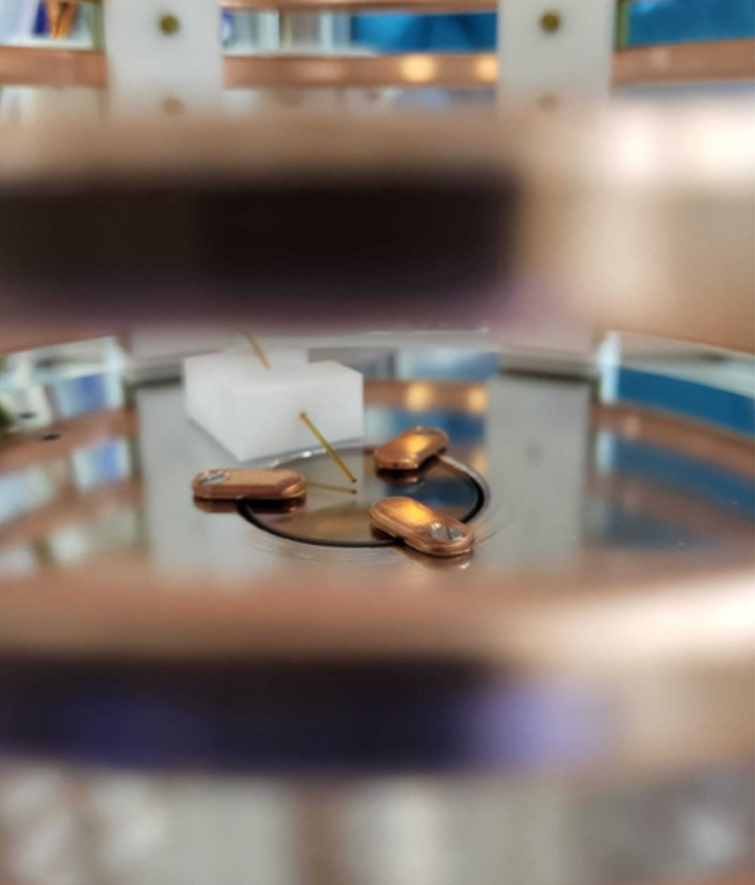} \end{minipage}
    \end{tabular}
        \caption{\textbf{(Left)} Full view of the assembly of the \SI{2.6}{m} TPC. \textbf{(Top right)} Top section of the TPC, displaying the anode and gate electrode meshes, as well the photosensor array facing the sensitive volume. \textbf{(Bottom right)} Photocathode plate fixed using three copper fasteners. An optical fibre is used to flash light from an external xenon lamp.  \label{fig:modules}}
    \end{figure}

The screws used to join the components of the modules are also made of PAI to avoid the presence of metallic components near the copper rings. Using data provided by the manufacturer~\cite{Ventura:torlon}, we estimate that the TPC contracts by approximately $\SI{0.4}{\percent}$ at LXe temperatures, which translates into a $\SI{\ca 10}{mm}$ reduction over the $\SI{2.6}{m}$ drift region. On the other hand, the weight of the field cage leads to an elongation, estimated to be about $\SI{6}{mm}$ using finite element simulations, partially compensated by the buoyancy of the immersed system. These changes in the length of the TPC due to contraction-expansion informed the design of the high-voltage system, as will be explained in \autoref{sec:hv}. \autoref{fig:modules} shows a complete view of the assembled field cage, along with other subsystems introduced below.

The top stack of the TPC is shown in the right top part of \autoref{fig:modules}, where the gate (bottom) and anode (top) electrodes are visible. These electrodes are constructed as hexagonal meshes welded to the centre of a \SI{4}{mm} wide circular frame. The vertical distance between the two rings is \SI{8}{mm}, resulting in a distance of \SI{12}{mm} between the meshes. Both electrodes are biased by a CAEN NDT1740 NIM high-voltage power supply module.

To produce electrons at the bottom of the TPC, a photocathode  is installed at the cathode plate. The photocathode was produced in-house by depositing $\SI{50}{\micro m}$ thick gold films on a \SI{2}{mm} thick fused-silica substrate disc following the same procedure as in ref.~\cite{Baudis:2023ywo}. The photocathode is illuminated with an external Hamamatsu L7685 xenon flash lamp, which produces UV-light pulses with wavelengths from 190 to \SI{2000}{nm}. These pulses are transmitted to the TPC using a LewVac solarisation-resistant fused-silica optical fibre with a polyimide buffer. The fibre is fixed and directed towards the centre of the photocathode using a PTFE holder, as seen in \autoref{fig:modules} (bottom right). To maintain the components at the same electrical potential, the plate is attached to the cathode using three copper fasteners.
    \subsection{The cathode high-voltage delivery system}
\label{sec:hv}
\begin{figure}[t]
    \centering
        \includegraphics[width=\textwidth]{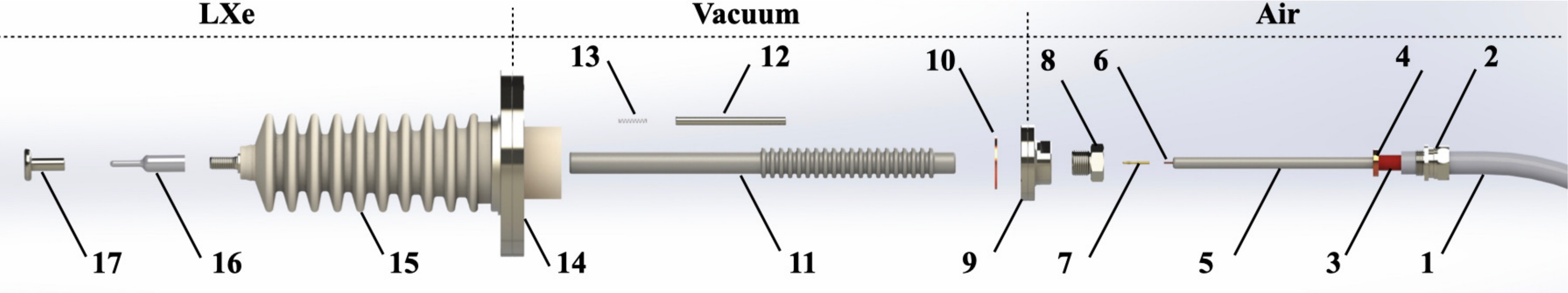}
    \caption{Exploded view of the cathode high-voltage delivery system components across the air, vacuum and LXe regions: (1) plated steel outer shield, (2) outer shield connector, (3) PVC cable outer layer, (4) copper ground shield, (5) HMWPE cable insulator, (6) copper inner conductor, (7) cable terminal, (8) grounding screw, (9) custom DN40CF flange, (10) DN40CF gasket, (11) HMWPE insulator, (12) stainless steel connector, (13) stainless steel spring, (14) DN125CF flange, (15) ceramic feedthrough, (16) round-edged high-voltage terminal, (17) cup-spring mechanism.}
    \label{fig:hv_assembly}
\end{figure}

\FloatBarrier
Current generation of LXe TPCs typically operate with electric drift fields within the range of \qtyrange{20}{200}{V/cm}~\cite{XENON:2024wpa, LZ:2019sgr}. In Xenoscope, the drift field is generated by establishing a potential difference between a stainless steel cathode plate and the gate electrode. 
The cathode plate is biased through a newly designed and custom-made HV delivery system inserted upward from the lower section of the outer vessel. The feedthrough consists of three components, shown in a horizontal exploded view in \autoref{fig:hv_assembly}: an air-to-vacuum feedthrough, a vacuum-to-LXe feedthrough and a cup-spring mechanism which delivers the high voltage to the cathode, located in the inner vessel. 

The HV power supply is a negative polarity Heinzinger PNC 100000 rated to \qty{100}{kV}. The voltage is transmitted from the power supply to the cryostat by means of a Heinzinger cable model HVC100 \cite{hvcable} rated for \qty{100}{kV} and composed of five layers: a copper inner conductor, a black high-molecular weight polyethylene (HMWPE) core insulator, a white HMWPE insulator, a copper shielding braid, and a polyvinyl chloride (PVC) sheath. The air-to-vacuum feedthrough consists of the HV cable cryofitted to a custom DN40CF flange attached to the bottom section of the outer cryostat vessel. The feedthrough is grounded by pinching the copper shield braid from the cable into the custom flange. The vacuum-to-LXe feedthrough extends the connection to the inner vessel by means of a commercial ceramic feedthrough from CeramTec \cite{CeramTec} rated to \qty{100}{kV}, \qty{4}{K}, and \qty{8.6}{bar}. The HV cable and the ceramic feedthrough are connected through a spring-loaded stainless steel connector housed inside a custom-made polyethylene connector. On the LXe side, the ceramic feedthrough features a round-edged pin that mechanically connects to the cup-spring system. This system comprises a round-edged stainless steel cylinder that houses a concave surface and a coil spring. At one end, the cylinder is threaded to a round-edged cup that is attached to the cathode plate, ensuring a reliable electrical connection. The opposite end of the cylinder makes contact with the pin from the ceramic feedthrough, establishing reliable mechanical and electrical connectivity. The design of the cup-spring system ensures a robust but flexible electrical contact with the cathode, accommodating potential thermal contractions at LXe temperatures, as detailed in \autoref{sec:cage}.

Following a successful helium leak test in a small-scale vacuum setup, which achieved a leak rate of \SI{e-9}{mbar\cdot l\per s}, the air-to-vacuum feedthrough and the vacuum-to-LXe feedthrough were assembled in the outer vessel, with a leak rate of \SI{e-8}{mbar\cdot l\per s}. Both feedthroughs were electrically conditioned by progressively increasing the applied voltage to improve their resistance to electrical breakdown. This controlled process involved intentionally inducing breakdowns to eliminate field emitters, which contribute to breakdown initiation \cite{osti_2204110}. The conditioning was performed by gradually increasing the voltage while monitoring possible discharges via voltage and current fluctuations in the HV power supply, as well as changes in the vacuum level of the cryostat. If a discharge occurred, the voltage was lowered until stable conditions were observed for 10-\qty{15}{minutes}. The voltage was then slowly increased to where the last discharge took place and maintained there for a few minutes before increasing it further. After several days of conditioning, a voltage of \qty{-50}{kV} was achieved, enabling drift fields of up to \SI{\ca 190}{V/cm} for the height of the Xenoscope TPC.

Temperature and pressure tests were simultaneously conducted to verify the performance of the ceramic feedthrough. The feedthrough was covered in liquid nitrogen to mimic the temperature conditions expected during filling cycles. The pressure conditions were achieved through the boil-off of the liquid nitrogen inside the inner vessel. The test resulted in stable behaviour at an overpressure of \qty{4}{bar} and a temperature of \qty{120}{K} for a duration of five minutes.

    \FloatBarrier
\subsection{Liquid level monitoring and control}
\label{sec:level}

\begin{figure}[t]
	\centering
	
	\sbox\twosubbox{%
		\resizebox{\dimexpr.99\textwidth-1em-1em-1em}{!}{%
			\includegraphics[trim={0 0 0 0},clip,height=3cm]{figures/levelmeters/slm_crop_scale}%
			\includegraphics[trim={0 0 0 0},clip,height=3cm]{figures/levelmeters/llm_overlap_scale}%
            \includegraphics[trim={0 0 0 0},clip,height=3cm]{figures/levelmeters/llm_ring_crop_scale}%
            \includegraphics[trim={0 0 0 0},clip,height=3cm]{figures/levelmeters/weir_upper_scale}%
		}%
	}
	\setlength{\twosubht}{\ht\twosubbox}
	
		\includegraphics[trim={0 0 0 0},clip,height=\twosubht]{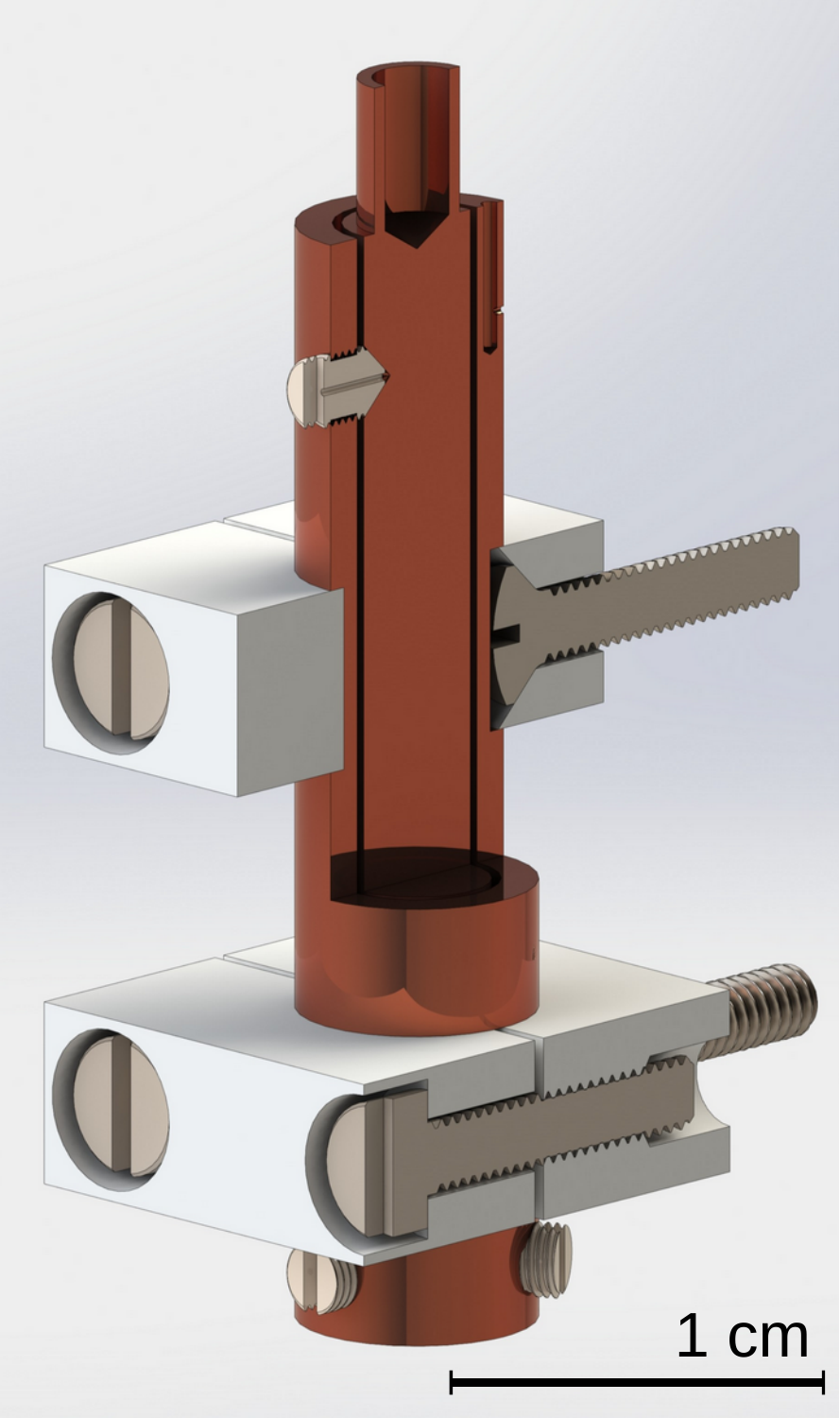}%
    \quad
		\includegraphics[trim={0 0 0 0},clip,height=\twosubht]{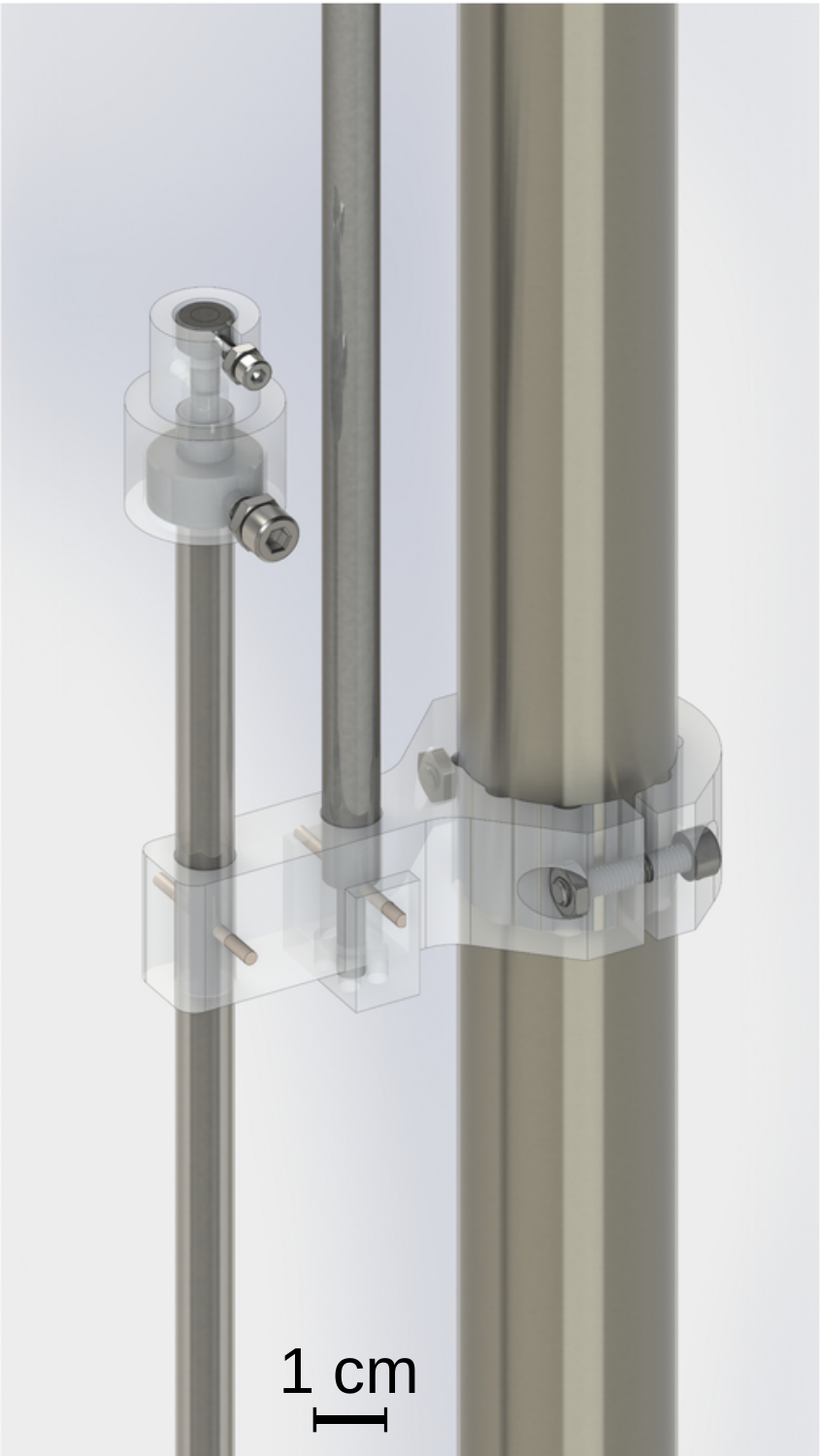}%
    \quad
		\includegraphics[trim={0 0 0 0},clip,height=\twosubht]{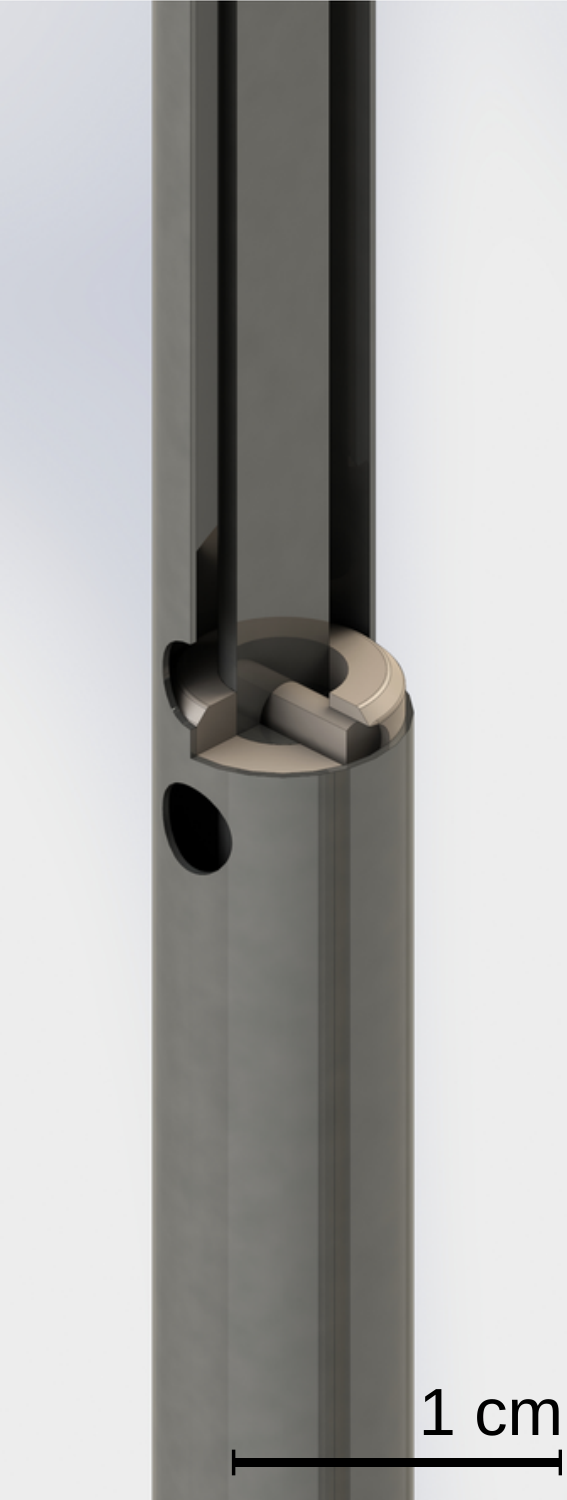}%
    \quad
		\includegraphics[trim={0 0 0 0},clip,height=\twosubht]{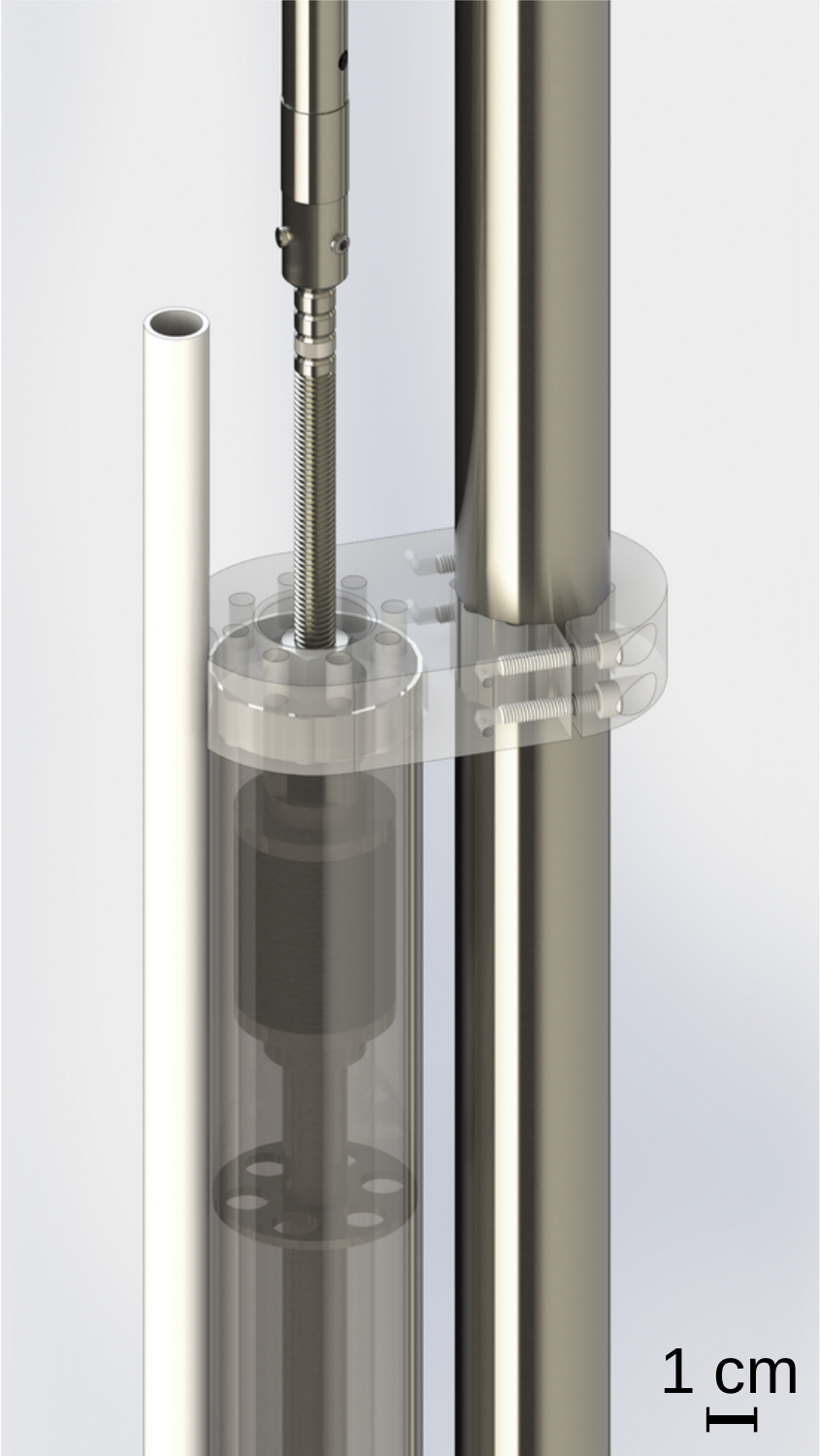}%
	
	\caption[Liquid level monitoring and control system.]{Computer-aided design (CAD) renderings of the individual sub-systems for the xenon liquid level monitoring and control. \textbf{(Left)}~Short level metre including its holder structure (cutaway view at two levels). \textbf{(Centre left)}~Overlapping region of the two long level metre segments attached to the support rod. \textbf{(Centre right)}~Close-up view of one of the spacer rings inside the long level metre with venting holes in the outer electrode cylinder (cutaway view). \textbf{(Right)}~Top part of the weir system mounted to the support rod.}
	\label{fig:leveling}
\end{figure}

To ensure adequate and stable electron extraction and amplification for the S2 signal generation, an untilted level of the xenon liquid-gas interface must be set and measured reliably. For this purpose, a liquid level monitoring and control system, illustrated in \autoref{fig:leveling}, was deployed in Xenoscope. The design improves upon the one implemented in the XENONnT detector
, achieving a more stable and reliable readout.

For the level measurement informing the xenon filling and recuperation processes, a \mbox{two-piece} long level metre (LLM) covers the full length of the TPC. A set of three short level metres (SLMs) provides more precise liquid level information between the anode and the gate electrode. The symmetric placement of the \SI{30}{mm} long SLMs at every other TPC PAI pillar, about \SI{9}{cm} from the central axis of the TPC, additionally enables tilt estimates. Both level metre types are based on a capacitive level readout, exploiting the differing relative permittivities $\varepsilon_r$ of the liquid ($\varepsilon_r = \SIrange[]{1.85}{1.96}{}$~\cite{Marcoux:1970die, Amey:2004die, Sawada:2003cap}) and gas ($\varepsilon_r \approx 1$) phases. 
With a cylindrical capacitor geometry at an inner diameter of the outer electrode of \SI{4.5}{mm} (\SI{7.5}{mm}) and an outer diameter of the inner electrode of \SI{4.0}{mm} (\SI{4.0}{mm}), the SLMs (LLM) experience an expected \SI{0.085}{pF/mm} (\SI{0.45}{pF/mm}) capacitance change per liquid level unit height. This allows for a conversion from the capacitance reading to the vertical position of the liquid-gas interface. If fully submerged in liquid, the SLMs (LLM) reach their expected maximum capacitance of \SI{27.8}{pF} (\SI{251}{pF}). 

The oxygen-free copper cylinders of the SLMs are aligned concentrically by two sets of three dielectric PAI alignment screws, whereas the stainless steel electrodes of each LLM segment are equipped with five equidistant Polyether ether ketone (PEEK) spacer rings. These rings serve additionally as calibration markers: at the height of these elements the liquid level inside the LLM does not change, resulting in an identifiable capacitance plateau, as illustrated in \autoref{fig:llm_calibration}. Furthermore, the rings are accompanied by venting holes in the outer cylinder for an unobstructed xenon flow.

The level metre capacitance values are read out with a custom readout board employing a smartec Universal Transducer Interface (UTI)~\cite{Vandergoes:1997uti, Smartec:2016dat} chip per level metre unit, capable of cancelling parasitic capacitances and auto-calibrating offset and gain. The maximum recommended readout capacitance of the UTI chips of \SI{300}{pF} enforces the segmentation of the LLM into two independent pieces of \SI{1440}{mm} length each for an increased allowed capacitance per unit level height and thus an improved resolution. The UTI chips are sequentially operated and read out by a PIC16F18857 microcontroller, which communicates to the readout computer via a galvanically isolated \mbox{LTC2863-2} RS485 transceiver for noise suppression.

The adjustable setting of the liquid level height and its decoupling from temporal external variables is realised through a weir system with a maximum capacity of about \qty{0.8}{l} of LXe. The weir consists of a vessel made of two concentric cylindrical walls, open at the top, with edge-welded belows in the centre cylinder. Compressing and expanding the bellows alters the vertical position of leak points from where the xenon flows into the outer overspill cylinder, effectively setting the height of the liquid-gas interface. A stainless steel actuation rod connects the leak tube part to a magnetically-coupled push-pull linear motion feedthrough, attached to a DN40CF vacuum feedthrough on the cryostat top flange and actuated by a remotely controllable 17HS-240E stepper motor. The LXe inside the weir is extracted at its bottom and drawn to the purification loop through a half-inch PFA tube.

The LLM and weir are both secured via PTFE mounting clamps to a hollow segmented stainless steel support rod alongside the off-centre TPC. Furthermore, the support rod accommodates modular, crescent-shaped PTFE fillers, visible in the bottom of \autoref{fig:modules} (right). They are meant to displace volume in the non-instrumented parts of the inner cryostat vessel and hence reduce the amount of xenon required during operation. The presently employed two stacks of three \SI{7}{cm} high filler units each result in a total displacement of \SI{\ca 16.2}{kg} of LXe.

    \subsection{Photosensor array}\label{sec:sipm}
\label{sec:toparray}

\begin{figure}[t]
    \centering
    \begin{subfigure}[c]{0.49\textwidth}
        \includegraphics[width=\textwidth,trim={0 -40 0 0},clip]{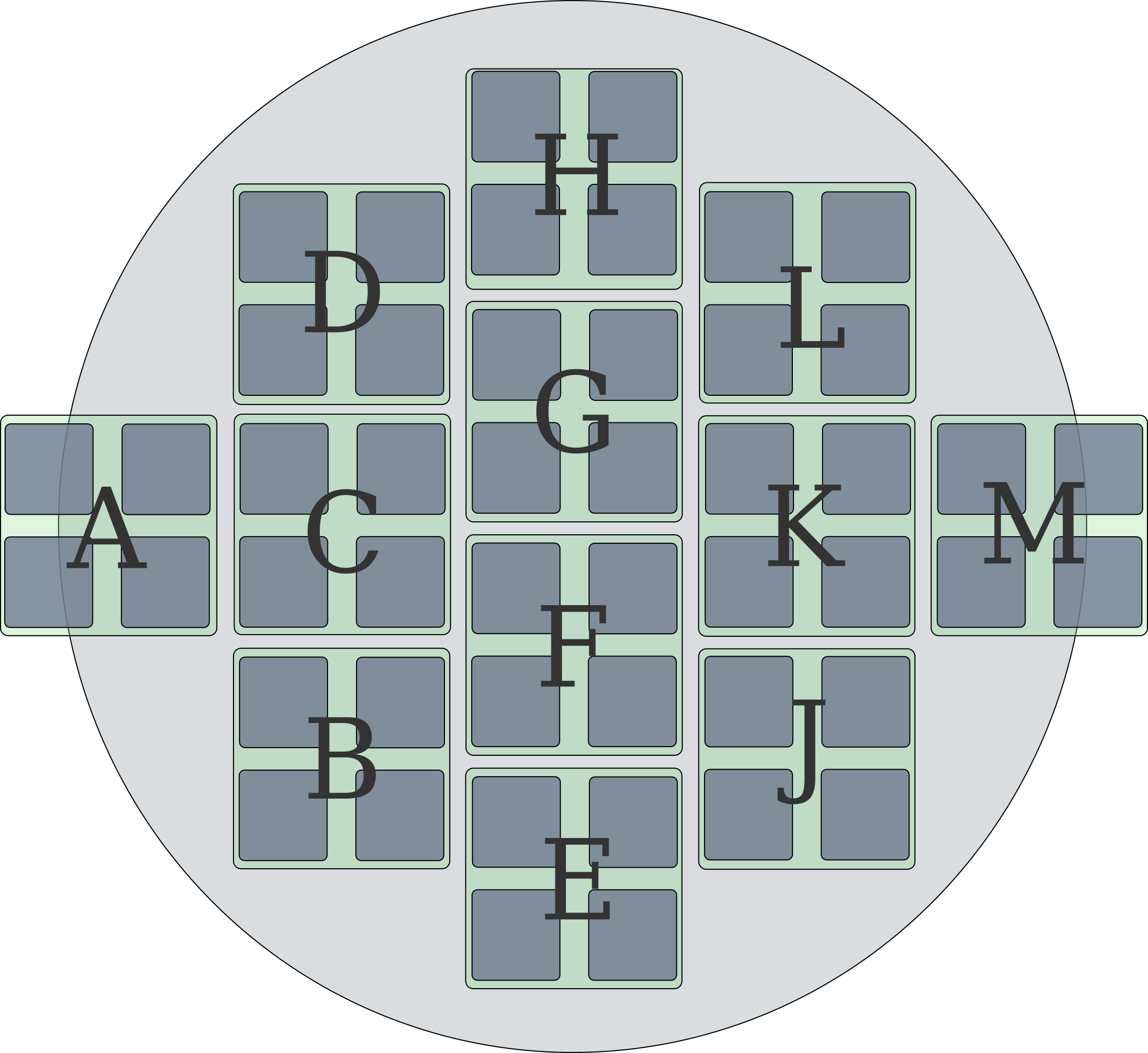}
    \end{subfigure}
    \hfill
    \begin{subfigure}[c]{0.49\textwidth}
        \includegraphics[width=\textwidth]{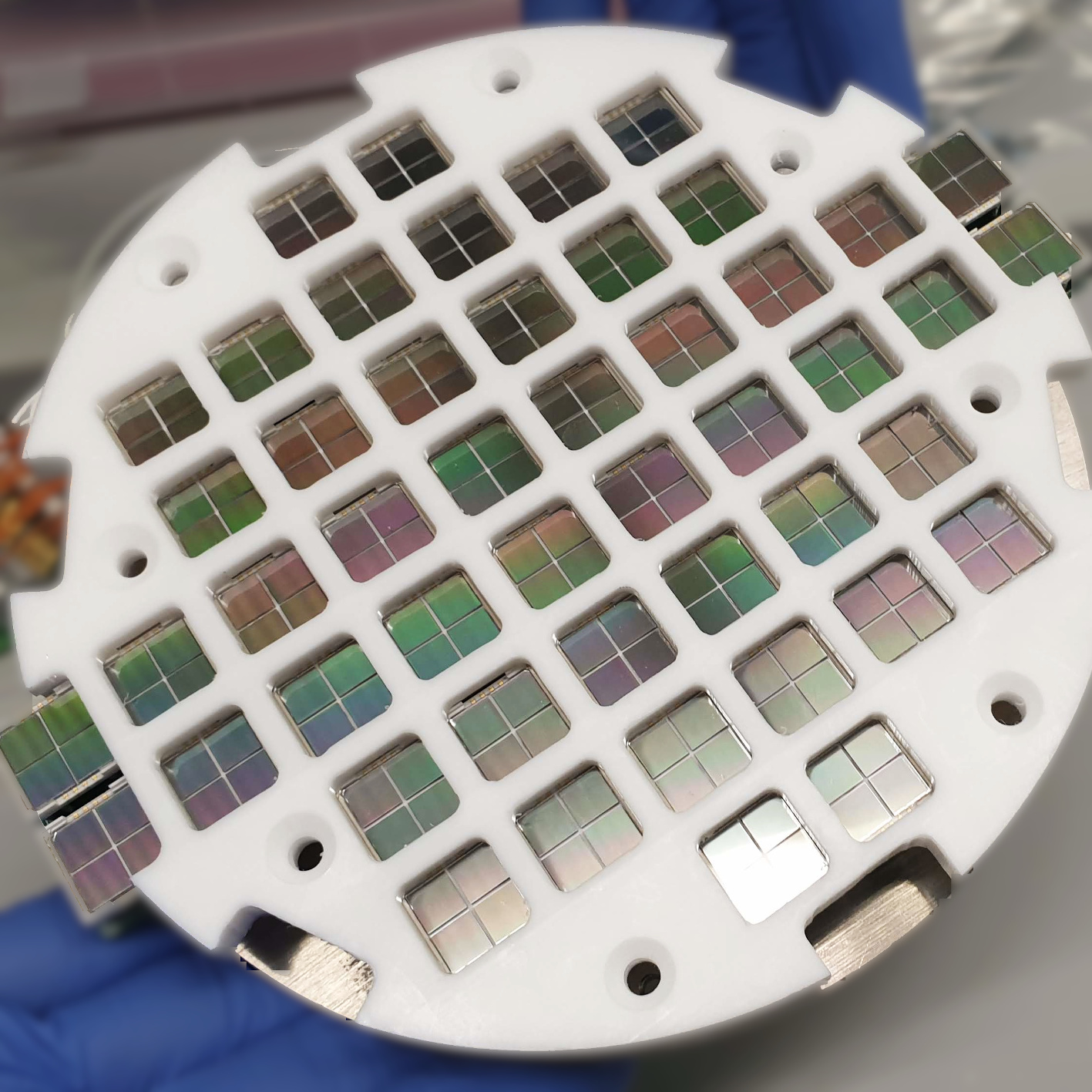}
        \label{fig:loadedarray}
    \end{subfigure}
    \vspace{3mm}
    \begin{subfigure}[c]{0.95\textwidth}
        \includegraphics[width=\textwidth]{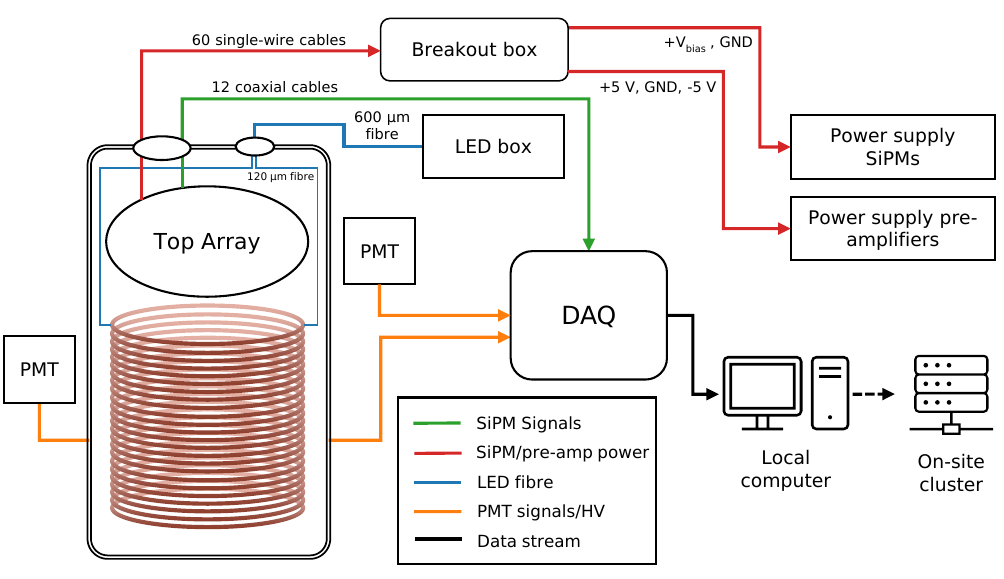}
        \label{fig:wiring}
    \end{subfigure}
    \caption{\textbf{(Top left)} Sketch of the tile distribution and naming in the top SiPM array. \textbf{(Top right)} Fully assembled top array before installation in the TPC. \textbf{(Bottom)} Schematic of the cabling routing of the SiPM array and the auxiliary coincidence trigger PMTs (described in \autoref{sec:commissioning}).} 
    \label{fig:fullarray}
\end{figure}

The Xenoscope TPC is instrumented by a large-area array of VUV-sensitive SiPMs designed for scalability through modular tiling of the photosensor units. The array holds up to \num{192} SiPM units of~\sbs. These are packaged in \numproduct{2 x 2} quad modules (S13371-6050CQ-02 units from Hamamatsu) with a total sensitive area of \tbt. The quads are grouped in sets of four for each readout PCB, named tiles, as detailed in~\cite{Peres:2023nbw}. The array hosts 12 tiles, named alphabetically from A to M, excluding the letter I, as sketched in \autoref{fig:fullarray}.
%
The tiles serve at the same time as holders for the SiPM units, voltage distributors, and pre-amplifiers for the signals. The readout scheme is based on the design proposed in~\cite{Arneodo:2017ozr}, where the amplified output is the analogue summed signal, optimised to exclude contributions from non-triggered SiPMs. Each pre-amplifying circuit is loaded with an OPA847 operational amplifier from Texas Instruments and provides an effective $\times 20$ amplification factor to the summed signal. Tiles B, G, and M are equipped with on-board PT100 resistance temperature detectors (RTDs) to monitor the temperature of the GXe in proximity to the photosensors.

The array of tiles is screwed to a stainless steel plate for mechanical stability with \qty{6}{\mm} PTFE standoffs to ensure the protection of the wiring on the backside of the PCB. When assembled at the top of the TPC, the array is secured by the stainless steel plate fitting into grooves in the PAI pillars with the photosensors facing downwards, as shown in \autoref{fig:modules} (left). The distance between the anode mesh and the optical plane is \qty{15}{\mm}. The plate measures \qty{160}{\milli\metre} in diameter, \qty{4}{\milli\metre} in thickness, and it spans the entirety of the cross-section of the cylindrical TPC. Rounded squares of \qty{27.5}{\milli\metre} length are cut from the plate where the back of the tiles are located, allowing wires to pass through without being damaged or excessively bent. At the side, six M2 screw-holes match the position of the PAI pillars to securely tighten the array in place with vented screws. To ensure that the SiPM units are secured in place and prevent their dislodging, a PTFE cover with cutouts is placed in front of the sensors and screwed to the stainless steel plate. The PTFE cover spans the full circular area of the TPC and extends \qty{3.8}{mm} below the quartz windows of the SiPMs. Although the edges of the holes cause shadow effects at large incident angles, these are not a major concern since the primary goal is to detect $\mathcal{O}(10^2$-$10^4)$\,photoelectrons (PEs) S2 signals.

Each of the twelve tiles requires five power supply wires (three for the pre-amplifier and two for the SiPMs) and one signal cable. Considering these requirements, a custom-made potted electrical feedthrough was commissioned from VACOM~\cite{vacom} to connect wires from the vacuum or xenon gas side to the air side. The potted feedthrough has 63 Kapton-insulated wires, 14 coax PTFE-jacketed cables and two HV-robust single Kapton-insulated wires (unused), and is mounted on a DN40CF flange assembled on the top of the setup. To facilitate both single-wire and coax cable connections on the inside, a set of custom PTFE connectors is used, following the design developed for XENON1T~\cite{Aprile:2012zx}. Outside of the cryostat, the signal cables are routed from the top of the facility to the lower floor, where the data acquisition (DAQ) rack is placed. There, the cables are secured in a patch panel with SMA socket connectors on the front, from where a connection is made to the digitisers. The \num{60} power wires of the top array are connected to a breakout box which merges all the wires at the same potential, inline with safety current limiters. From the breakout box, five wires are routed to the power supplies: the pre-amplifiers are powered by a standard current-limited power supply, while the SiPM bias voltage is supplied by a custom low-noise power supply.

In order to monitor the gains of the photosensors during operation, a triggered calibration system is set up using a blue light-emitting diode (LED). A KA-4040QBS-D LED from Kingbright is installed on the outside of the vessel inside a small dark box, pointed at an SMA 905 connector. From the LED box connector, a \qty{600}{\micro\metre} optical fibre guides the light to a UV/VIS dual-fibre feedthrough from Accu-Glass~\cite{accuglass}. The feedthrough is mounted on a DN40CF flange on the top of the assembly and connects to the fibre through another SMA 905 connector. The second connection on the feedthrough is designed to host the fibre coming from the xenon flash lamp. The LED light is split into three \qty{120}{\micro\metre} bare fibres routed in the inner volume of the detector alongside the PAI pillars. These are wrapped around the top part of the field cage and secured in PTFE holders attached to the pillars, pointing upward towards the top array. One of the bare LED fibres is partially shown in~\autoref{fig:modules} (upper right). In the final assembly, two out of the three fibres were verified to be working, situated close to tile A and tile M. A pulse generator sends synchronously a signal to power the LED inside the box with a user-defined shape and amplitude, and a transistor–transistor logic (TTL) pulse to the data acquisition to serve as a trigger.

    \section{Commissioning of the TPC}
\label{sec:commissioning}

The commissioning run of the \qty{2.6}{m} dual-phase TPC of Xenoscope lasted from late April to early August 2024. During this run, the gas purifying getter was not available due to maintenance, limiting the scope of the work to test and benchmark each sub-system required for future measurements with the setup: liquid level setting and monitoring, HV delivery, and light readout with the SiPM photosensor array. 

As a result of the lack of purification, the impurity concentration in the xenon gas was expected to severely reduce the ability to drift charges through the xenon bulk, therefore preventing the formation of S2 signals from interactions far below the gate electrode. Furthermore, to avoid permanent damage to the TPC in the event of dielectric breakdown, the target drift field was set to \qty{100}{V/cm}, corresponding to an absolute cathode voltage below \qty{30}{kV}. After the end of the commissioning work reported here, the maintenance of the getter was concluded and xenon purification will be available in future runs.

\subsection{Xenon filling and levelling}

Prior to the start of the xenon filling, the system was thoroughly leak-checked and outgassed through vacuum pumping. The bulk of the xenon was then transferred from the storage array to the cryostat during \qty{\ca 2}{weeks}, aided by circulating the xenon at a flow of about \qty{50}{slpm} and by operating the pre-cooler system described in~\cite{Baudis:2023ywo}. In total, \qty{\ca 360}{kg} of xenon were filled until the liquid level reached the anode. During the filling, the mass inside the cryostat was monitored via the difference in the integrated flows of the flow metre, which monitors the inflow of xenon to the inner vessel, and the flow controller, which sets the  recirculation speed of xenon from the inner vessel into the purification loop.

\begin{figure}[t]
    \sbox\twosubbox{%
		\resizebox{\dimexpr.99\textwidth-1em}{!}{%
			\includegraphics[trim={6 0 0 -2},clip,height=3cm]{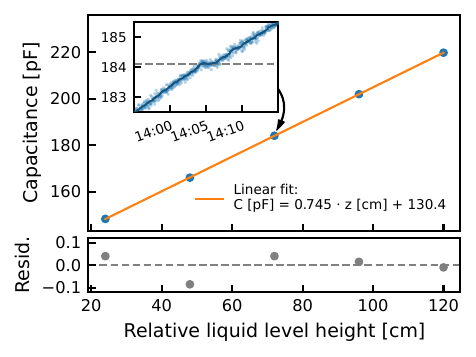}%
			\includegraphics[trim={0 0 6 0},clip,height=3cm]{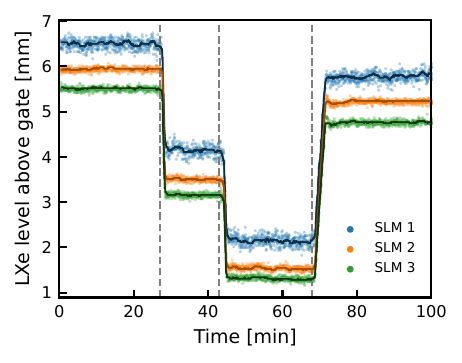}%
		}%
	}
	\setlength{\twosubht}{\ht\twosubbox}
    \centering
		\includegraphics[trim={6 0 0 -2},clip,height=0.98\twosubht]{figures/levelmeters/llm_calibration_fit_lower_inset_small.pdf}%
    \quad
		\includegraphics[trim={0 0 6 0},clip,height=0.98\twosubht]{figures/levelmeters/weir_tests_20240723_xminutes_small.pdf}%
    \caption{\textbf{(Left)} Calibration of the lower long-level metre segment. The capacitance plateau values at the equidistant spacer rings are identified, as illustrated in the inset for the third spacer ring, and fitted linearly to the known vertical ring positions. \textbf{(Right)} Weir leak point height adjustments as monitored by the three SLMs. The first two dashed lines mark a lowering by \SI{\ca 2}{mm} each, whereas the third one indicates a rise by \SI{\ca 4}{mm}. The offset discrepancy between the readings of the individual SLMs is mainly attributable to systematic uncertainties of the calibration, but could also partially arise from a possible tilt of the TPC.}  
    \label{fig:llm_calibration}
\end{figure}

After the first \qty{\ca 40}{kg}, the rising of the liquid level was observed with the LLM readout. The two-stage level metre system was calibrated with the aid of the spacer rings, as discussed in \autoref{sec:level} and illustrated in \autoref{fig:llm_calibration} (left). Derived estimates for the filled xenon mass inside the cryostat were consistent with values inferred from the integrated filling flow.
To calibrate the SLMs, the LXe level was temporarily raised such that these sensors were fully submerged, thus providing an estimate on their maximum capacitance $C_{\mathrm{max}}$. 
The reading after the instantaneous capacitance increase from capillarity once the liquid first touched the lower end of the level metre cylinders defines the value $C_{\mathrm{cap}}$.
Together with the capacitance in gas $C_{\mathrm{min}}$, the relative liquid height conversion from each sensor is obtained as $z(C) = \SI{30}{mm}\cdot(C-C_{\mathrm{cap}})/(C_{\mathrm{max}}-C_{\mathrm{min}})$. With these corrected level estimates at hand and the height-adjustable weir, whose functionality is demonstrated in \autoref{fig:llm_calibration} (right), a liquid level centred between the anode and gate electrode, corresponding to \qty{\ca 6}{mm} above the gate, was set.

\subsection{High voltage setting}

Once the liquid level was set, the voltage applied to the electrodes was ramped up. The complete ramp-up operation spanned two weeks with intervals for data taking with the SiPMs. 

First, the same voltage with opposite polarity was set on the anode and gate electrodes. This configuration minimises the risk of discharges to the grounded elements closer to the electrodes, such as the SiPM array and SLMs. During the ramp up, the CAEN power supply had a single trip incident. Without any noticeable effect to any other sub-system, the top stack voltage was successfully set once more and remained stable until the end of the commissioning run. The final set of voltages was \qty{3.6}{kV} at the anode and \qty{-3.6}{kV} at the gate, inducing an electric extraction field of \qty{6.7}{kV\per cm} (\qty{\ca 3.4}{kV/cm}) in the gas (liquid) phase. 

The cathode voltage was similarly ramped up in parallel to avoid any downward accelerating electrons hitting the photocathode. The cathode voltage was then slowly ramped to \qty{-26}{kV}, defining a drift field of \qty{\ca 86}{V/cm}  over the full \qty{2.6}{m} TPC. After four days of data acquired, the cathode voltage was set to \qty{-29}{kV}, achieving a drift field of \SI{\ca 100}{V/cm} over the full \qty{2.6}{m} TPC. The cathode voltage was stable for seven days at this target configuration until the end of the run and subsequent ramp-down of the voltage of the electrodes.

\subsection{SiPM calibration and gain monitoring}

From the beginning of the run, the SiPMs were turned on regularly. Their signals were recorded by two CAEN v1724 digitisers~\cite{v1724} communicating via daisy-chain and connected to a computer. The raw data files in ROOT format were copied to a local server for redundancy and processing. Further data handling and processing was performed with PyLArS~\cite{pylars}, an open-source custom-made Python framework designed for SiPM characterisation and later adapted for peak processing of a multi-channel SiPM readout.

\begin{figure}[t]
    \centering
    \begin{subfigure}[c]{0.5\textwidth}
        \includegraphics[width=\textwidth]{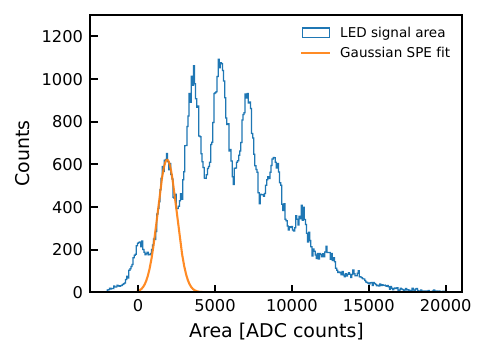}
    \end{subfigure}%
    \hfill
    \begin{subfigure}[c]{0.5\textwidth}
        \includegraphics[width=\textwidth]{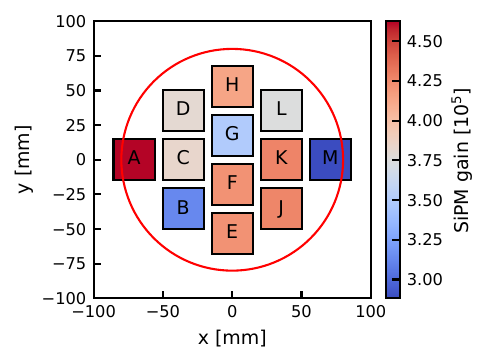}
    \end{subfigure}%
    \vfill
    \begin{subfigure}[c]{1\textwidth}
        \includegraphics[width = 1\textwidth]{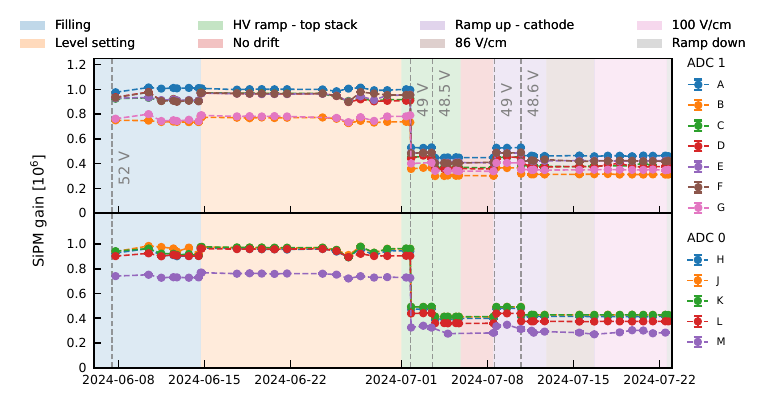}
    \end{subfigure}%
    \caption{\textbf{(Upper left)} Charge spectrum of an LED calibration run for a highly-illuminated SiPM channel. The orange line corresponds to the Gaussian fit to the SPE peak. \textbf{(Upper right)} Map of the gains of the different tiles in the top array with a $\mathrm{V_{bias}}$ of \qty{48.6}{V}. The red circle denotes the \qty{16}{cm} inner radius of the field cage. Three channels (B, G, and M) consistently showed a lower gain throughout the whole run. \textbf{(Lower)} Measured gain of the SiPM tiles (error bars within data markers). The channels are shown in two panels with the ones connected to ADC\#1 on the top and the ones connected to ADC\#0 on the bottom. Vertical dashed gray lines indicate changes in the bias voltage of the SiPMs. The coloured bands correspond to the different stages of the commissioning described in the text.} 
    \label{fig:gain_calculation}
\end{figure}

The gain of each tile was monitored with regular LED calibrations. For this purpose, the digitisers were directly triggered by the pulse generator TTL synchronisation pulse and \qty{3}{\micro\second} waveforms were recorded. Data was acquired for five different LED intensities each time to populate the single photo-electron (SPE) region of both high and low illuminated tiles. The charge was reconstructed by integrating a \qty{400}{ns} window, encompassing the full LED signal. A resulting charge spectrum is shown in \autoref{fig:gain_calculation}~(upper left), where the SPE charge peak is fitted with a Gaussian function. Its mean value is used to calculate the gain of the photosensor.

The occupancy of each tile, defined as the mean number of PE detected for each LED light level, was also calculated from the charge spectra and monitored over time. Processing of the LED data occurred automatically after its upload to the server with the functions available in PyLArs, providing a map of the gains of all the channels, as shown in \autoref{fig:gain_calculation}~(upper right).

The bias voltage of the SiPMs was adjusted throughout the commissioning to avoid saturation of the signals, especially noticeable once the extraction field was established and the first S2s were observed. Apart from these operational changes, the SiPM gains remained stable, with a maximum relative standard deviation of the gains per tile below \qty{3.5}{\percent} during stable operation periods. \autoref{fig:gain_calculation}~(bottom) details the gain evolution of all the tiles over the whole operational period. The achieved gains of the SiPMs are in-line with the standard photosensor operation in current large liquid xenon detectors, such as XENONnT, LZ or PandaX-4T. Three tiles --- B, G, and M --- consistently showed a lower gain throughout the run. As these were the ones where on-board PT100s were placed, the lower gain can be likely associated with a slightly higher operation temperature due to heat dissipation in the RTD. As the gains were overall stable, this effect can be corrected for and does not impact the performance of the array.

\subsection{Event acquisition in dual-phase mode}
The signals from physical interactions were recorded in both muon-coincidence, and channel trigger mode. To trigger on and record signals attributable to cosmic muons, two BC-412 plastic scintillator panels were placed on opposite sides of the outer vessel with \qty{\ca 30}{cm} vertical separation, corresponding to an incident angle of \qty{\ca 10}{\deg}. The panels were positioned close to the top of the cryostat, with the path of the incoming cosmic muons crossing the liquid-gas boundary and the very top of the liquid phase. The scintillators were respectively observed by two \num{3}-inch PMTs. Each PMT signal is duplicated in a CAEN Fan in/Fan out module and routed to both a CAEN N840 leading-edge discriminator and one of the CAEN v1724 digitisers. The discriminator was set with a threshold above the electronic noise level for both PMT signals. The resulting NIM pulse is read by a CAEN N455 coincidence logic unit set to AND mode, providing the coincidence trigger. The trigger is supplied to the digitisers for data acquisition in ``muon-coincidence mode'' and to an ORTEC 997 analogue counter for easier monitoring of the number of coincidence events acquired.

To record signals independently of the muon coincidence trigger with a rate manageable for the DAQ ($<$\qty{1}{MHz}), an analogue channel trigger was implemented. The signals of three centre tiles (C, F, and G, see \autoref{fig:fullarray}, left), were multiplied to three outputs in a CAEN N625 linear Fan in/Fan out module. The first set of outputs is supplied to the digitisers in order to record the signal. The second set of signals is read by an oscilloscope to allow monitoring during data-taking and to facilitate the debugging of the setup. The final set of signals is routed to a leading-edge discriminator with a defined threshold of \qty{15}{mV}. The resulting NIM pulse was then routed to the CAEN N455 logic unit where a triple coincidence scheme is set up. Finally, the output pulse was supplied to both the digitisers and the oscilloscope to act as trigger. 

\begin{figure}[t]
    \centering
    \includegraphics[width = 1.\textwidth]{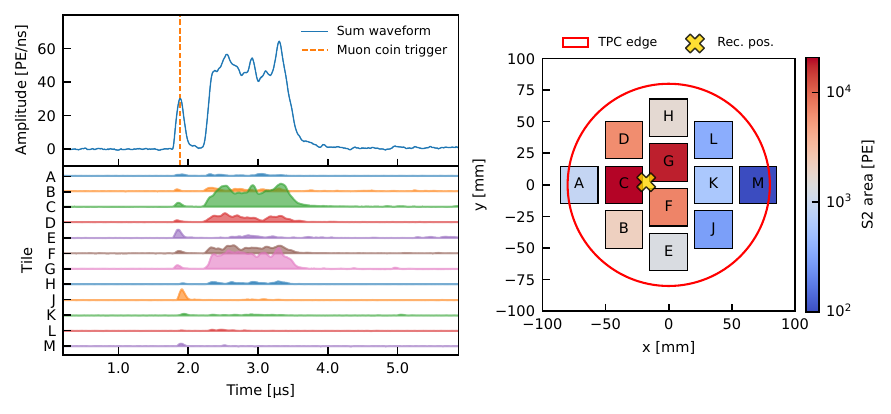}
    \caption{\textbf{(Left)} Waveform of an S1-S2 event from a muon crossing the TPC in the LXe region around the gate. The sharp S1 signal at \qty{\ca 2}{\us} corresponds to the prompt light emission upon interaction of a muon with the xenon target, while the second and wider S2 signal corresponds to the electroluminescence light produced by the charges drifted from the interaction site to the xenon gas region. The top panel shows the summed waveform and trigger signal from the muon-coincidence system and the bottom panel the individual waveforms of each recorded channel (to scale). The total reconstructed areas of the S1 and S2 signals are \qty{4.2e3}{PE} and \qty{5.96e4}{PE}, respectively. \textbf{(Right)} Observed hit pattern of the same event. The reconstructed position of the interaction vertex in the x-y plane is marked with a cross, as calculated with a centre of gravity algorithm based on the relative area of the S2 signal in each tile. The red circle delimits the \qty{16}{cm} inner diameter of the field cage.}
    \label{fig:s1_s2}
\end{figure}

After gaseous xenon was transferred to the vessel, prompt scintillation light signals (S1-like) were observed. These display a typical duration of up to \qty{1}{\us} and rise-time of \qty{\ca 0.1}{\us}. After filling with LXe and establishing the extraction field, S2-like events started to be observed. These are characterised by a wider pulse, with a duration of 1 to \qty{6}{\us}, rise-time above \qty{\ca 0.2}{\us}, and clear localisation on the top array. They were identified as electroluminescence signals from interactions within the gas phase or in the liquid phase between the gate electrode and the liquid-gas boundary. Three main topologies of signals were observed: S1-like signals with no following correlated S2-like signal, S2-like wide signals with no preceding S1-like signal, and signals with an initial S1-like peak and a following S2-like peak with separations of \qtyrange{0}{10}{\us}. Signals with S1-S2 separation longer than $>$\qty{10}{\us}, corresponding to events from deeper in the TPC, were not observed, as, given the lack of xenon purification, the drifting charges were absorbed by impurities before reaching the extraction region. Correlated S1-S2 events were also observed in muon-coincidence mode, where the trigger signal from the PMTs of the muon panels was coincident with the observed S1 signal, followed by an S2 signal. An example of these events is shown in \autoref{fig:s1_s2}. The observation of such events from cosmic muons confirms the proper functioning state of the dual-phase TPC and all its directly related subsystems. 

After S1-S2 events from the top were observed, the plastic scintillator panels were moved to the lowest position to record data for cosmic muons traversing the TPC at the height of the cathode. In this configuration, S1 signals were still observed by the SiPM array in correlation with the scintillator trigger. 

    \section{Summary and outlook}
\label{sec:outlook}

We report on the design, installation and commissioning of a \qty{2.6}{m} tall dual-phase LXe TPC in the Xenoscope facility at the University of Zurich. This is the tallest xenon TPC in operation to date, and it is part of the rich R\&D programme of the DARWIN and XLZD collaborations, aimed at constructing a large, next-generation LXe observatory.

We described the main TPC subsystems such as the field cage, the custom-made high-voltage delivery system, the liquid level monitoring and control system, and the SiPM photosensor array. During the commissioning reported in this manuscript, we filled the cryostat with \qty{\ca 360}{kg} of LXe, monitoring the process with the segmented long level metre. The weir and the short level metres were used to adjust the final level and assess the horizontal alignment of the TPC.
We set a drift field of \SI{\ca100}{V/cm} and an extraction field of \SI{6}{kV/cm} by biasing the cathode, gate and anode at \qty{-29}{kV}, \qty{-3.6}{kV} and \qty{3.6}{kV}, respectively. 
The photosensors were characterised with periodic LED calibrations during the entire run, showing stable gains with a maximum relative standard deviation of \qty{3.5}{\percent}. To complete the commissioning phase, we acquired data by tagging muons crossing the TPC via two PMTs coupled to plastic scintillators placed on opposite sides at the top of the cryostat. We selected triple coincidence events between the muon-trigger system and the SiPM array. Despite the fact that xenon purification was not available during this work and thus the detector was not optimised to detect and reconstruct events in the whole volume of the TPC, we observed correlated scintillation and ionisation signals from cosmic muon interactions near the top of the detector, validating the dual-phase TPC working mode.

The commissioning of Xenoscope and its \qty{2.6}{m} high TPC is one of the planned steps towards the scale up of the detector size required for the next generation of LXe direct dark matter experiments. After concluding the maintenance on the xenon purification getter, further data taking is scheduled for 2025 to demonstrate electron drift along the whole depth of the TPC and perform electron cloud diffusion measurements. Following this goal, and within the timeline to inform the design of XLZD, we foresee an upgrade of the systems to measure LXe optical properties on a large scale by 2027, such as light attenuation, and of the SiPM array to increase the granularity of the photosensor readout plane.

\acknowledgments

This work was supported by the European Research Council (ERC) under the European Union’s Horizon 2020 research and innovation programme, grant agreement No. 742789 (Xenoscope), by the SNF grant 20FL20-216573, by the SNF grant 200020-219290, as well as by the European Union’s Horizon 2020 research and innovation programme under the Marie Skłodowska-Curie grant agreement No 860881-HIDDeN. We thank the mechanical and electronics workshops in the Physics Department for their continuous support. We thank J. M. Disdier for the provided support and advice.

\appendix
\section{Gas phase purification}
\label{sec:appendix}

The GXe phase was not recirculated neither during the facility commissioning~\cite{Baudis:2021ipf} nor in the purity monitor run~\cite{Baudis:2023ywo}. Instead, the extraction line for the purification loop was connected to the LXe phase of the detector. This purification loop relies purely on the transfer of impurities from the GXe to the LXe phase through the condensation of xenon. This transfer process is slow, inefficient, and negatively impacts the LXe purity.

\begin{figure}[htb]
    \centering
    \includegraphics[width=\textwidth]{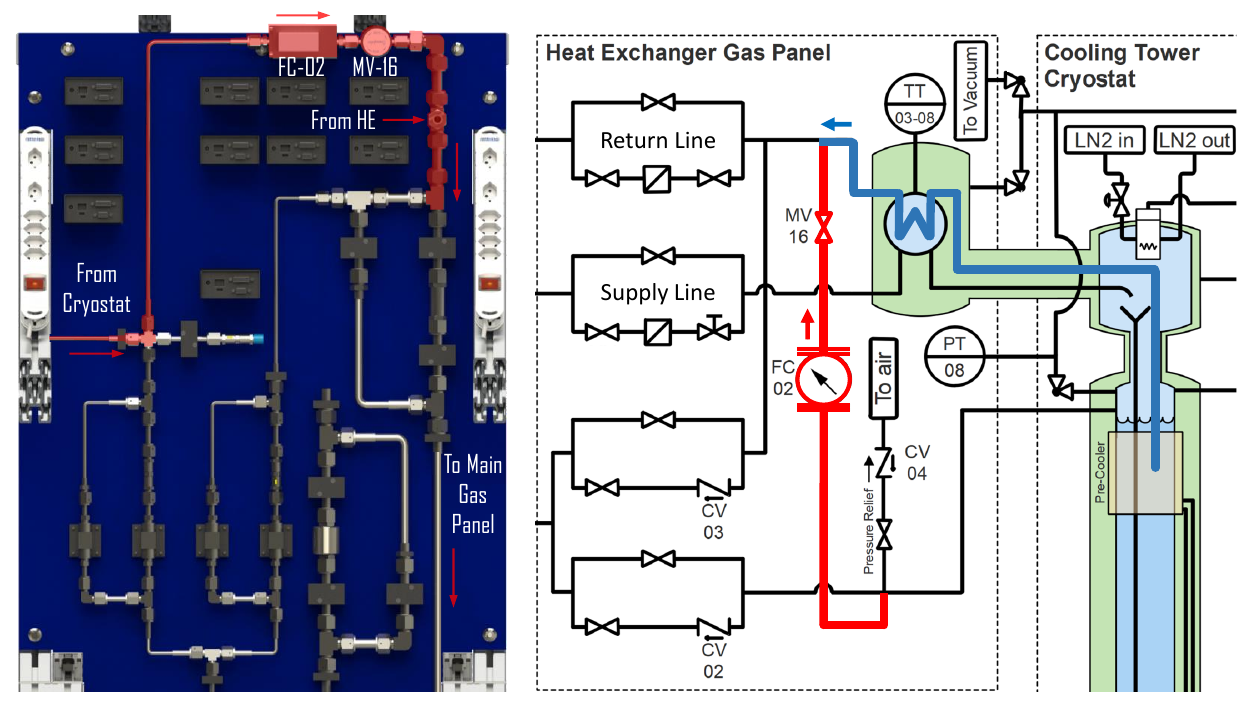}
    \caption{Gas-phase purification line (red) shown in the schematic of the heat-exchanger gas panel at the top of Xenoscope \textbf{(left)} and in the the piping and instrumentation diagram \textbf{(right)}. This line allows for direct GXe phase extraction, bypassing the LXe extraction and heat exchanger path (light blue). A mass flow controller (FC-02) controls the flow out of the gas-phase, while the main flow controller (FC-01) controls the total flow in the purification line. Valve MV-16 can be used as a shut-off valve. Adapted from~\cite{girard_thesis}.}
    \label{fig:gas_line}
\end{figure}

Adding a gas extraction line parallel to the liquid extraction line of the heat exchanger enables the purification of both gaseous and liquid xenon phases within a single purification loop. Figure~\ref{fig:gas_line} presents a schematic of this line as integrated into the filtration and safety gas panel (left), along with its layout in the piping and instrumentation diagram (right; see~\cite{Baudis:2021ipf} for the entire system description). The pressure relief line, connected to a gas feedthrough on the top flange, is now also bridged to the other end to the return line, bypassing the heat exchanger (red path). A flow controller (FC-02, with range from 0 to \SI{20}{slpm}) regulates the mass flow of the recirculated GXe, while the overall mass flow is still controlled by the flow controller FC-01. The newly added valve MV-16 can be used as a shut-off valve as needed. This setup introduces additional heat load on the cryostat when operative, as all the xenon routed back to the cryostat via the supply line comes into direct contact with the outflow from the heat exchanger. However, this extra heat load is compensated by the cooling tower.

The gas-phase purification line was installed and tested right after the purity monitor run and successfully leak-checked to a level better than \qty{5e-9}{mbar\cdot l\per\second}. Subsequently, the system was filled with \SI{2.0}{bar} of GXe and recirculated for \SI{\ca 12}{h} with the mass flow controller FC-01 set to \SI{50.0}{slpm}. The mass flow controller FC-02 was cycled from \SI{0.0}{slpm} to \SI{3.0}{slpm} to confirm its functionality. The flow in FC-01 stayed nominal throughout the test, validating the GXe purification line concept. The line was then utilized during the TPC commissioning described in this work, where a stable dual-phase configuration with a steady LXe level was achieved while recirculating xenon from both, the gas and the liquid phase.

\bibliographystyle{jhep}
\bibliography{main}  
\end{document}